# All-Altermagnetic Tunnel Junction of $RuO_2/NiF_2/RuO_2$

Long Zhang (张龙)[1], Guangxin Ni (倪广鑫)[2,3,*] , Xuehao Wu (吴学镐)[4], Guoying Gao (高国营)[1,*]

[1]School of Physics and Wuhan National High Magnetic Field Center, Huazhong University of Science and Technology, Wuhan 430074, People's Republic of China

[2]Department of Physics, Florida State University, Tallahassee, Florida 32306, USA

[3]National High Magnetic Field Laboratory, Tallahassee, Florida 32310, USA

[4]Department of Physics, Columbia University, New York, NY 10027, USA

[*]Corresponding authors. Email: guangxin.ni@magnet.fsu.edu; guoying_gao@mail.hust.edu.cn

Emerging altermagnets with zero net magnetic moment and moment-dependent spin splitting offer a promising avenue for antiferromagnetic spintronic devices, yet their integration into magnetic tunnel junctions has been hindered by reliance on ferromagnetic electrodes (introducing stray fields) or by limited functionality (non-tunable magnetoresistance without spin filtering). Here, we propose an all-altermagnetic tunnel junction (AAMTJ) paradigm composed exclusively of altermagnets—exemplified by experimentally feasible $RuO_2/NiF_2/RuO_2$. By introducing an altermagnetic $NiF_2$ barrier, the achieved tunneling magnetoresistances of 11,704%, 2,496% and 1,892% for $RuO_2/NiF_2/RuO_2$ are much higher than that of 221% for $RuO_2/TiO_2/RuO_2$ with a nonmagnetic $TiO_2$ barrier. High spin filtering efficiencies of ~90% are also obtained. This architecture unlocks multistate high magnetoresistance and spin filtering via magnetization control of the electrodes and barrier, stemming from their synergistic and antagonistic alignments of momentum-dependent altermagnetic spin-splitting. Importantly, high tunneling magnetoresistances are still achieved in the AAMTJ with $TiO_2$ spacer of $RuO_2/TiO_2/NiF_2/TiO_2/RuO_2$. Our AAMTJ inherently exhibits low consumption and zero stray field, with nonrelativistic spin splitting and vanishing magnetic moment, combining the advantages of both ferromagnetic and antiferromagnetic tunnel junctions. This AAMTJ paradigm opens an interesting avenue within the area of high-performance altermagnet-based tunnel junctions.

*1. Introduction.* Altermagnets [1,2] represent an emerging class of collinear magnetic systems characterized by a vanishing net magnetic moment and momentum-dependent nonrelativistic spin splitting with *d*-/*g*-/*i*-wave spin-momentum interactions. Altermagnetism arises from specific spin and crystal symmetries with time-reversal-breaking magneto-responses.[3-5] Several altermagnetic (AM) materials including $NiF_2$,[6] MnTe,[7] CrSb,[8] and $KV_2Se_2O$,[9] have been experimentally realized. Spin-splitting switching in $RuO_2$,[10,11] CrSb,[8] and $Mn_5Si_3$[12] has been demonstrated experimentally, confirming their potential for controlling spin-polarized currents. Altermagnet-based devices and switching phenomena have been demonstrated.[13-15] By transcending the conventional dichotomy of antiferromagnets and ferromagnets,[16-21] altermagnets, possessing immunity to stray-field interference, fast spin dynamics, high nonvolatility, and low consumption, hold significant promise for applications in magnetic memory and logic devices.[22-26]

In magnetic random-access memory, conventional magnetic tunnel junctions (MTJs) typically comprise two ferromagnetic (FM) electrodes with high spin polarization, separated by a non-magnetic semiconducting or insulating barrier.[27,28] In conventional MTJs, the tunneling magnetoresistance (TMR) arises from the matching or mismatching of spin-polarized conduction channels in the two electrodes, which is governed by the relative spin orientations.[29] The spin-filtering effect in spin-filtering MTJs refers to the phenomenon in which a magnetic barrier presents different barrier heights for different spin channels.[30] Altermagnets have been incorporated into MTJs in three primary configurations: (1) An AM metal serving as one electrode, paired with a FM half-metal (HM) or near-HM electrode (e.g., $RuO_2/TiO_2/CrO_2$[31] and $CrSb/In_2Se_3/Fe_2GaTe_2$[32]); (2) An AM semiconductor or insulator acting as the barrier layer, combined with a FM HM electrode (e.g., $IrO_2/MnF_2/CrO_2$[33]); (3) Two AM layers separated by a non-magnetic barrier (e.g., $RuO_2/TiO_2/RuO_2$[34] and $Ag/V_2Te_2O/BiOCl/V_2Te_2O/Ag$[35]), where TMR stems from the relative orientation of AM spin splitting.

We here propose an all-altermagnetic tunnel junction (AAMTJ) model, schematically illustrated in Fig. 1(a), where both electrodes and the barrier layer consist entirely of AM materials. This design eliminates dependence on FM HMs, thereby removing stray fields and enabling fast spin dynamics. The AM magnetization orientations of the two AM electrodes and the AM barrier provide multiple controllable degrees of freedom. Switching the AM spin-splitting is anticipated to generate substantial TMR and spin-filtering effect, offering high nonvolatility and low power consumption. Experimentally realized altermagnets, $RuO_2$ (metal)[36] and $NiF_2$ (insulator),[6] are selected to construct the AAMTJ, serving as electrodes and barrier, respectively. Both $RuO_2$ and $NiF_2$ share a rutile structure with excellent lattice matching, facilitating high-quality heterostructures fabrication.

In this Letter, we investigate a $RuO_2/NiF_2/RuO_2$ AAMTJ using the nonequilibrium Green's function (NEGF) approach combined with density functional theory (DFT). By controlling the magnetization alignments of the electrode and barrier, the AM spin splitting can be flipped, producing unexpected TMR of 11,704%, which is much higher than that of 221% for the non-AAMTJ of $RuO_2/TiO_2/RuO_2$. This AAMTJ also exhibits widely tunable TMRs ranging from 30% to 11,704%, and high and low spin-filtering efficiencies, establishing it as a compelling platform for exploring multi-controlled AM spin transport with zero stray field, high nonvolatility, and low energy consumption.

*2. Computational Method.* First-principles calculations are performed using DFT within the Vienna ab initio Simulation Package (VASP).[37] The Perdew-Burke-Ernzerhof (PBE) exchange-correlation functional of the generalized gradient approximation (GGA)[38] is employed, incorporating a Hubbard $U$.[39] According to earlier studies, the $U_{\mathrm{eff}}$ is set to 4 eV for Ni-3$d$ orbitals[40] and 2 eV for Ru-4$d$ orbitals.[41,42] A plane-wave cutoff energy of 600 eV is utilized, with convergence criteria of $10^{-6}$ eV for energy and 0.001 eV/Å for force. Monkhorst-Pack $k$-meshes of $7 \times 7 \times 14$ and $10 \times 10 \times 14$ are utilized for bulk $RuO_2$ and $NiF_2$ with (110) and (001) crystal faces, respectively. Using DFT combined with the NEGF approach, spin transport properties of $RuO_2/NiF_2/RuO_2$ AAMTJ are calculated in the QuantumWise Atomistix Toolkit (ATK) package.[43] A cutoff energy of 150 Hartree and $k$-meshes $14 \times 7 \times 150$ and $150 \times 150$ are used for self-consistent and transmission calculations, respectively. The tunneling magnetoresistance (TMR) is defined as $\mathrm{TMR} = (T_{\mathrm{P}} - T_{\mathrm{AP}})/T_{\mathrm{AP}} \times 100\%$, where $T_P$ and $T_{AP}$ are transmission coefficients in the spin parallel and antiparallel alignments between two altermagnets. The spin-filtering efficiency $\eta$ is attained from $\eta = (T_{\uparrow} - T_{\downarrow})/(T_{\uparrow} + T_{\downarrow}) \times 100\%$, where $T_{\uparrow}$ and $T_{\downarrow}$ are transmission coefficients for spin-up and spin-down channels, respectively. All materials and AAMTJs are fully relaxed. Similar computational methods and convergence tests have been applied in our previous works.[32,44,45]

*3. Results and Discussion.* The rutile-type $NiF_2$ and $RuO_2$ structures exhibit centro-symmetry, with magnetic Ni and Ru atoms serving as their inversion centers, respectively. Both $NiF_2$ and $RuO_2$ crystallize in the $P4_2/mnm$ space group. Cleaving planes based on the (110) and (001) crystal faces yields two bulk structures as displayed in Figs. 2(a) and 2(b). Two similar bulk structures have been presented previously.[31,34] Spin-resolved charge densities [Figs. 2(c) and 2(d)] demonstrate their AM configurations, with two opposite magnetic sublattices, the Ni atoms in $NiF_2$ and the Ru atoms in $RuO_2$ exhibit atomic magnetic moments of ±1.78 and ±1.16 $\mu_B$, respectively. Their magnetic space group is $P4'_2/mnm'$, violating $TP\tau$ and $U\tau$ symmetries, where $T$, $P$, $\tau$, and $U$ stand for the time reversal, spatial inversion, half lattice translation, and spinor symmetry, respectively. Both materials display *d*-wave symmetry, with their opposite magnetic sublattices related by a 90° planar rotation ($C_{4z}$).

The high-symmetry points within the Brillouin zone are illustrated in Fig. 2(e). As shown in Figs. 3(a)-3(d), $NiF_2$ and $RuO_2$ possess insulating (band gap of 4.29 eV) and metallic features, respectively. Our calculated spin-resolved electronic structures are consistent with previous reports.[31,40] The broken $PT$ symmetries of $NiF_2$ and $RuO_2$ allow for momentum-dependent and nonrelativistic spin splitting, forming symmetry-connected spin-momentum locking. This effect arises from exchange coupling rather than spin-orbit coupling (SOC). Pronounced spin splitting appears along the Γ-X path for the bulk structure with (110) crystal face and along the M-Γ path for the bulk structure with (001) crystal face, suggesting that the [110] transport direction can yield spin polarization. In contrast, the spin-resolved bands remain degenerate along the Γ-Z path, indicating the absence of spin discrimination for transport along the [001] direction. This orientation-dependent nature of AM spin transport has been verified previously.[31,33] Thus, the [110] transport direction is the focus herein. The orbital behaviors of the dominant atoms near the Fermi level further explicate the electronic and spin properties. Along the Γ-X path, the valence band maximum (VBM) of $NiF_2$ arises mainly from spin-up Ni-$d_{yz}$ and spin-down Ni-$d_{xz}$ orbitals, while the conduction band minimum (CBM) is dominated by Ni-$d_{x^2-y^2}$ and Ni-$d_{z^2}$ orbitals in both spin channels [Fig. S1(a)]. Similarly, the spin splitting near the Fermi level in $RuO_2$ is primarily governed by spin-up Ru-$d_{xz}$ and spin-down Ru-$d_{yz}$ orbitals [Fig. S1(b)].

Considering the structural, electronic, and magnetic properties of $NiF_2$ and $RuO_2$, together with their experimentally confirmed altermagnetism,[6,36] the $RuO_2/NiF_2/RuO_2$ AAMTJ is constructed along the [110] direction [Fig. 4(a)]. Bulk rutile $NiF_2/RuO_2$ with (110) crystal face possesses lattice constants of $a$ = $b$ = 4.65/4.55 × $\sqrt{2}$ = 6.58/6.43 Å and $c$ = 3.08/3.14 Å, consistent with reported values.[40,46] The lattice mismatch is calculated as $\Delta = (a_1\text{-}a_2)/a_2 \times 100\%$, in which $a_1$ and $a_2$ stand for the lattice constants of the two materials. The resulting lattice mismatch of $c$ and $a$ in our AAMTJ is only 1.7% and 2.2%, respectively, which is smaller than the experimentally demonstrated 6.5% in $Fe/MgAlO_x/Fe_4N$[47] and 4.3% in $Co_2MnSi/MgO/Co_2MnSi$,[48] confirming its structural feasibility. Two possible magnetization orientations are considered for both the $NiF_2$ barrier ($M_{2\text{-}1}/M_{2\text{-}2}$) and the right $RuO_2$ ($M_{3\text{-}1}/M_{3\text{-}2}$) electrode. Consequently, four distinct magnetization configurations are examined in this AAMTJ [Fig. 4(b)].

A high TMR ratio indicates a pronounced resistance contrast between data bits (“0” and “1”), while perfect spin-filtering efficiency transmits electrons selectively and efficiently, improving TMR and data fidelity. Both characteristics are crucial for high-performance and nonvolatile magnetic memory devices. Spin- and $\vec{k}_{//}$-transmission spectra of the $RuO_2/NiF_2/RuO_2$ AAMTJ in energy and momentum space [Figs. 4(c)-4(l)] are schematically illustrated in Fig. 1(b). The calculated values of $T_\uparrow$, $T_\downarrow$, $\eta$, and TMR are listed in Tables 1 and 2, which include six combinations of P and AP magnetization alignments. $T_{\mathrm{P}} = \sum_{\vec{k}_{//}} T_{\mathrm{P}}(\vec{k}_{//}) / N_k$ and $T_{\mathrm{AP}} = \sum_{\vec{k}_{//}} T_{\mathrm{AP}}(\vec{k}_{//}) / N_k$ denote the transmission coefficients for parallel (P) and antiparallel (AP) magnetization alignments, respectively. In-plane wave vector $\vec{k}_{//} = (k_x, k_y)$ is perpendicular to the transport direction.

For State-1, the spin-up transmission is predominantly localized along the Γ-Y line [Fig. 4(e)], while the spin-down transmission is distributed between the Γ-X and Y-M paths [Fig. 4(f)]. The dominance of the spin-up channel results in a high $\eta$ (93%). In contrast, in State-2, the transmission for both channels is concentrated near the Γ point and along the Γ-Y line [Figs. 4(g) and 4(h)], yielding comparable transmission coefficients and a low $\eta$ (-19%). Similarly, pronounced transmission along the Γ-Y line is observed for the spin-up (spin-down) channel in State-3 (State-4), resulting in a high $\eta$ value of 85% (-90%). A positive (negative) $\eta$ indicates preferential tunneling of spin-up (spin-down) electrons. The ability to selectively filter different spin channels under distinct magnetization states highlights the tunability and versatility of spin transport in the AAMTJ, which is advantageous for practical spintronic applications.

These transport phenomena can be further understood from spin-resolved band structures along the [110] transport direction. $NiF_2$ exhibits dominant spin-up character at both the VBM and CBM along the Γ-X path [Fig. 3(a)]. Similarly, $RuO_2$ shows predominant spin-up bands near the Fermi level along this direction [Fig. 3(b)]. This spin character facilitates efficient spin-up electron tunneling, leading to the large $T_{\uparrow}$ observed in State-1. For State-2, the magnetization reversal of the right $RuO_2$ electrode shifts its dominant near-Fermi-level bands to the spin-down channel, producing a strong mismatch with the spin-up-favored left electrode and barrier. This misalignment severely suppresses tunneling in both spin channels, leading to a giant TMR of 11,704% between State-1(P) vs State-2(AP) (considering State-1 and State-2 as P and AP alignments, respectively). This value surpasses most reported altermagnet-based junctions, such as 1,056% of $CrSb/In_2Se_3/Fe_2GaTe_2$[32] and 574% of $Ag/V_2Te_2O/BiOCl/V_2Te_2O/Ag$.[35] Such a giant TMR reflects the large difference between high- and low-resistance states, thereby reducing read/write error rates and improving device reliability. Even under equilibrium conditions ($E$=0) with the same magnetization reversal of the right $RuO_2$ electrode, the 11,704% TMR of the AAMTJ far exceeds 221% in $RuO_2/TiO_2/RuO_2$ (Fig. S2) (this value is consistent with reported 200%).[34] This enhancement originates from the AM $NiF_2$ barrier, which selectively promotes spin-up tunneling and suppresses spin-down tunneling owing to spin-differential contribution along the transport Γ-X direction [Fig. 3(a)]. Consequently, the AM $NiF_2$ barrier amplifies transport contrast between the P and AP states in $RuO_2/NiF_2/RuO_2$, in stark contrast to non-magnetic $TiO_2$ barrier with nearly identical tunneling for both spin channels in $RuO_2/TiO_2/RuO_2$.

State-3 differs from State-1 by a magnetization reversal of $NiF_2$ barrier, which moderately suppresses transmission in both spin channels. Nevertheless, since both $RuO_2$ electrodes continue to favor spin-up transport, $T_\uparrow$ in State-3 remains larger than $T_\downarrow$. The combination between State-1(P) and State-3(AP) yields a large TMR (2,496%), which exceeds the 150-170% reported for $NiF_2$-based junctions employing non-magnetic electrodes and barriers.[40] This result highlights the superiority of AM electrodes in our AAMTJ, which provide strong spin-polarized currents owing to the spin-up dominated bands near the Fermi level along the [110] direction in $RuO_2$ [Fig. 3(b)].

The spin-splitting flip in the conductive $RuO_2$ electrodes exerts a more pronounced impact on transmission than that of the wide-gap $NiF_2$ barrier. While $NiF_2$ primarily mediates tunneling, $RuO_2$ electrodes directly determine the available conducting states at the Fermi level.

Consequently, the total transmission in State-3 is higher than in State-2, leading to a TMR of 355% for State-3(P) vs State-2(AP). This finding demonstrates that, in an AAMTJ, spin switching within AM electrodes has a stronger effect on transport modulation than magnetization reversal within the AM barrier.

In State-4, the magnetizations of both the $NiF_2$ barrier and the right $RuO_2$ electrode are flipped relative to State-1. The resulting spin-splitting mismatch reduces the overall transmission, yielding a TMR of 1,892% between State-1(P) and State-4(AP). Tunneling is more efficient when the barrier and right electrode share the same dominant spin channel (spin-down in State-4) than when they are opposite, as in State-2. This alignment leads to a higher total transmission in State-4 than in State-2, producing a TMR of 493% between State-4(P) and State-2(AP). For State-4(P) vs State-3(AP), the spin-splitting of $NiF_2$ barrier is mismatched with the left $RuO_2$ electrode, suppressing tunneling regardless of the right electrode's alignments, yielding a low TMR (30%). The large variation in TMR from 11,704% to 30% achieved by flipping magnetizations of electrode and barrier highlights the AAMTJ's multistate functionality and its potential for practical controllability.

The interface non-idealities are important for tunneling properties. (1) In this AAMTJ, the tunneling magnetoresistance and spin-filtering effects arise from the spin splitting of the electrode and barrier. The difference in tunneling conductance among the various magnetic configurations results from a combination of the density of states at the Fermi level and symmetry matching. In the presence of interfacial intermixing, point defects, or interface roughness, short-range disorder at the interface primarily affects the injection intensity of tunneling electrons. However, as long as the crystallographic symmetry is preserved within the $RuO_2$ and $NiF_2$ layers, the differential decay rates for wave functions of different symmetries persist. Consequently, the relative conductance differences among the four magnetization states are expected to remain. (2) It was reported that in CoFeB/MgO, despite B diffusion and oxidation at the interface, a tunneling magnetoresistance of ~200% is still observed.[49] By analogy, for the $RuO_2/NiF_2$ system, if a thin intermixed layer (like Ni-Ru-O) forms at the interface, spin-polarized transport may be moderately affected but is unlikely to be completely suppressed. (3) Experimentally, post-annealing processes can be employed to optimize interfacial crystallinity. Furthermore, high-resolution scanning transmission electron microscopy and electron energy loss spectroscopy can be used to characterize the interface quality.

Spin-resolved projected local density of states (PLDOS) of the device [Figs. 5(a)-5(h)] and spin- and layer-resolved projected device density of states (PDDOS) [Figs. 6(a)-6(d)] further corroborate these interpretations derived from spin-resolved band structures and transport calculations. Device DOS is not a simple concatenation of the bulk DOS of the electrodes and barrier. Instead, it is computed directly from the Green's function of the entire open system using the NEGF method. Device DOS primarily focuses on the central region, which includes the electrode extended regions and the barrier layer, but does not directly include the electrodes themselves. The electrodes are semi-infinite periodic structures, while the electrode extended regions are geometrically identical to the electrode units. They serve as transition regions that ensure the electronic states from the electrodes can enter the central region smoothly and continuously, avoiding unphysical boundary effects caused by interface truncation. The PLDOS in Fig. 5 is the function of energy and position of the device, which is spatially resolved in continuous space of the device, and one can see from it the interface, local states and barrier. The layer-resolved PDDOS in Fig. 6 is the function of energy and layer ($RuO_2$, $NiF_2$) of the device, and one can see from it the transport contribution from each layer. In our calculation, the source of contribution for the device DOS is set to the left $RuO_2$ electrode. The left electrode extended region receives some electronic states injected from the left electrode. These states undergo exponential decay upon entering the $NiF_2$ barrier. By the time the electrons reach the right electrode extended region after tunneling through the barrier, their wavefunction amplitudes are substantially attenuated. Consequently, the electronic states detectable in the right electrode extended region are significantly fewer than those in the left electrode extended region (Fig. 6). Physical origin of the finite DOS in the $NiF_2$ barrier at the Fermi level: (1) The metallic wavefunctions from the left $RuO_2$ electrode extended region decay exponentially into the $NiF_2$ barrier, inducing a finite density of states within the barrier's band gap. (2) Orbital hybridization between $RuO_2$ and $NiF_2$ at the interface can lead to the formation of interface resonance states. Additionally, quantum confinement effects within the $NiF_2$ barrier may discretize energy levels, potentially introducing states near Fermi level that are absent in the bulk band gap. These effects collectively contribute to the non-zero DOS observed inside the $NiF_2$ barrier near the Fermi level (Fig. 6).

The PLDOS in Fig. 5 is the device DOS projected onto the transport direction. Within the NEGF framework, the electronic states in the extension region are calculated self-consistently in coupling with the barrier, and are influenced by multiple factors, such as interfacial coupling with the $NiF_2$ barrier, quantum confinement effects, and modulation by the magnetic state (Néel vector orientation) of the barrier. Therefore, the variation of the PLDOS in the left electrode extension with different magnetic configurations shown in Fig. 5 does not imply that the intrinsic properties of the left electrode itself have changed. Instead, it reflects the selectivity of electronic state channels incident from the left that can effectively participate in tunneling. When the magnetic configuration of the $NiF_2$ barrier changes, its decay rates for electron waves with different symmetries and wavevectors are altered, leading to a re-selection of the electronic states in the left electrode extension that are allowed to couple into the barrier. The contribution of these selected states to the PLDOS correspondingly varies. The PLDOS in the left electrode extension shown in Fig. 5 represents the joint response of the electrode coupled to the barrier, rather than the isolated intrinsic properties of the electrode. The PLDOS in the electrode extension region reflects which electronic states can effectively couple into the barrier. The $NiF_2$ barrier contacts the $RuO_2$ electrode extension region, leading to interface effects. When electrons are incident from the left electrode, they first enter the extension region between the electrode and the barrier. Here, the electron wave functions need to match the decaying modes in the barrier to enable tunneling. The $NiF_2$ barrier acts as a "filter." When the magnetization configuration of $NiF_2$ is altered, its attenuation of electrons with different wave vectors varies. Quantitatively, at the Fermi level, for the $RuO_2$ unit cell in the left electrode extension region in contact with $NiF_2$, the DOS values in State-1 for spin-up and spin-down directions are 0.096 and 0.031 $eV^{-1}$, respectively (Fig. 6), with a ratio of 3.131. After the magnetic moments of $NiF_2$ are flipped, i.e., in State-3, the DOS values for the two spin directions become 0.674 and 0.363 $eV^{-1}$, respectively, with a ratio of 1.857. These values quantitatively reflect the barrier-induced decay selectivity in the electrode extension.

For State-1 [Figs. 5(a), 5(b), and 6(a)], the predominance of spin-up electron tunneling is confirmed by a pronounced spin-up PDDOS peak near the Fermi level within the right electrode region, exceeding the spin-down counterpart. This demonstrates efficient spin-down filtering. The overall electronic states in State-1 are more than those in State-2, State-3, and State-4, validating the origin of the high TMR when State-1 serves as the P configuration. In contrast, State-2 exhibits diminished electronic states for both spin channels near the Fermi level [Figs. 5(c), 5(d), and 6(b)] due to mismatched transport channels between the electrodes, consistent with its poor spin-filtering efficiency. Conversely, State-3 and State-4 display a distinct predominance of spin-up and spin-down states [Figs. 5(e), 5(h), 6(c), and 6(d)], respectively, with majority-spin densities not only exceeding the opposite spin channel but also surpassing those in either spin channel of State-2. This observation accounts for the highly spin-polarized tunneling in State-3 and State-4, and the moderate TMRs achieved when these states are configured as P against State-2 as AP.

Finally, it is worth pointing out that the $RuO_2/NiF_2/RuO_2$ AAMTJ shows critical limitations. For example, (1) the current spacer-free structure suffers from the defect of being unable to switch independently due to strong interlayer coupling; (2) there is an inherent inconsistency in the Néel vectors for $NiF_2$ and $RuO_2$. Consequently, this specific structure is unrealistic for actual devices. The current model is employed primarily to explore the underlying physics of transport properties, temporarily without considering these structural and magnetic mismatch issues. So, by adding the $TiO_2$ spacer to the AAMTJ, we further evaluate the transport properties of the $RuO_2/TiO_2/NiF_2/TiO_2/RuO_2$ tunnel junction in Fig. S3 and Tables S1 and S2. It's found that TMRs of 28,091%, 6843% and 6390% and near-perfect spin filtering are still achieved. The transport behaviors in $RuO_2/TiO_2/NiF_2/TiO_2/RuO_2$ are qualitatively consistent with those in $RuO_2/NiF_2/RuO_2$, and thus future practical devices could be designed using similar spacer-inserted architectures.

*4. Conclusion.* In summary, we have proposed an AAMTJ architecture composed exclusively of AM systems [Fig. 1(a)], exemplified by $RuO_2$ electrodes and a $NiF_2$ barrier. With zero net magnetic moment and nonrelativistic spin splitting across the entire device, the AAMTJ combines the advantages of ferromagnets (spin polarization and high controllability) and antiferromagnets (zero stray field and fast spin dynamics), while avoiding their respective drawbacks. The AAMTJ architecture unlocks multistate spin transport, accompanied by widely tunable TMR and spin filtering achieved by flipping AM spin splitting through controlled magnetization of electrodes and barrier [Fig. 1(b)]. Within this fully AM platform, giant tunneling TMRs of 11,704%, 2,496% and 1,892%, and high spin-filtering efficiency of ~90% are realized in $RuO_2/NiF_2/RuO_2$. The TMRs are much higher than those of 221% for the non-AAMTJ of $RuO_2/TiO_2/RuO_2$. These effects arise from the synergistic and antagonistic spin-splitting alignments between AM electrodes and barrier. Our AAMTJ will stimulate subsequent theoretical and experimental explorations into AM tunnel junctions.

*Acknowledgments.* Guoying Gao acknowledges support from the National Natural Science Foundation of China (Grant No. 12174127). Guangxin Ni acknowledges support from the U.S. Department of Energy (DE-SC0022022), the U.S. National Science Foundation (DMR-2145074), and the ACS Petroleum Research Grant (PRF# 66465-DNI10).

# Figures and Tables

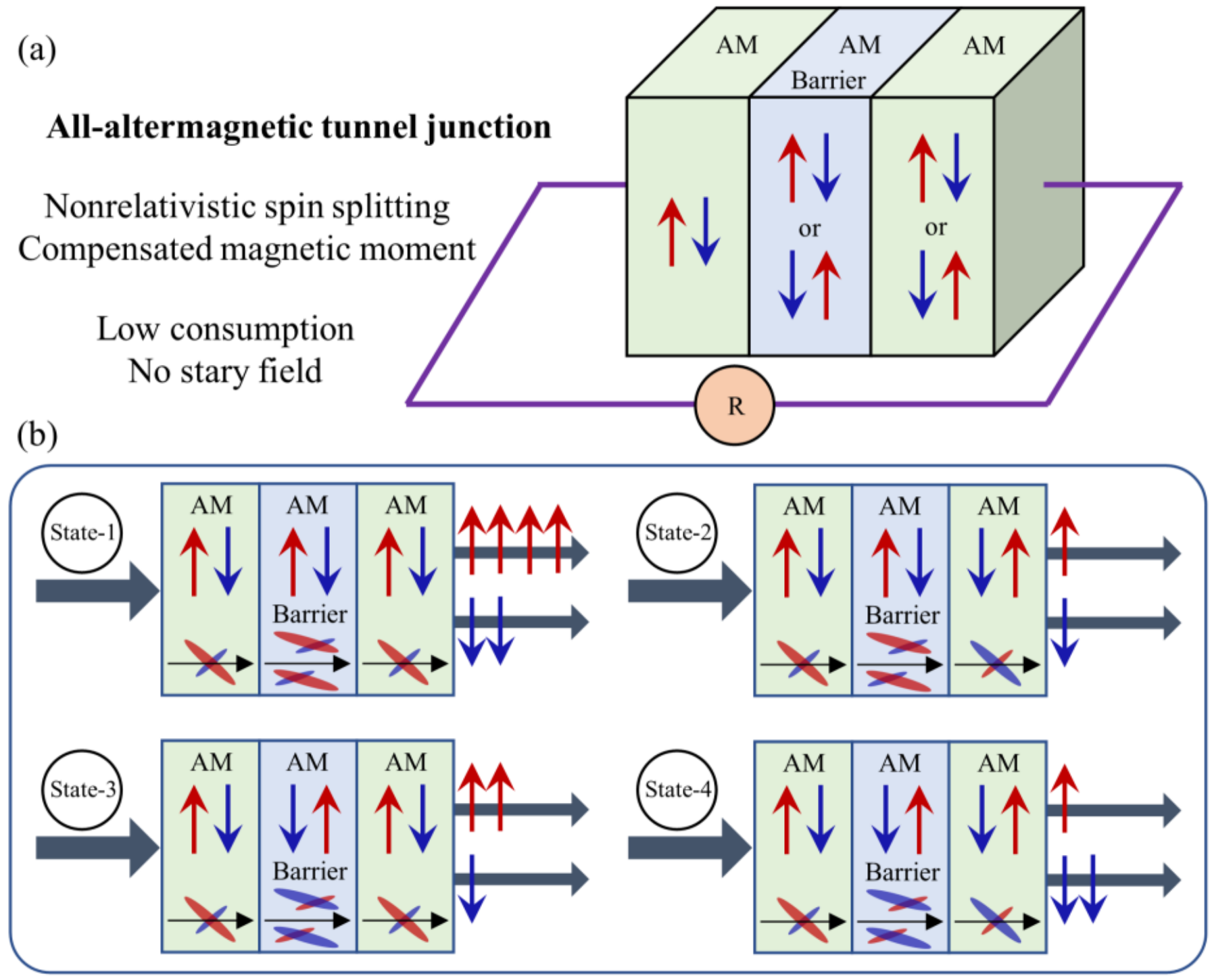

**Fig. 1.** Schematic illustrations of the all-altermagnetic tunnel junction (AAMTJ) with varying magnetization configurations (a) and multistate spin transport (b).

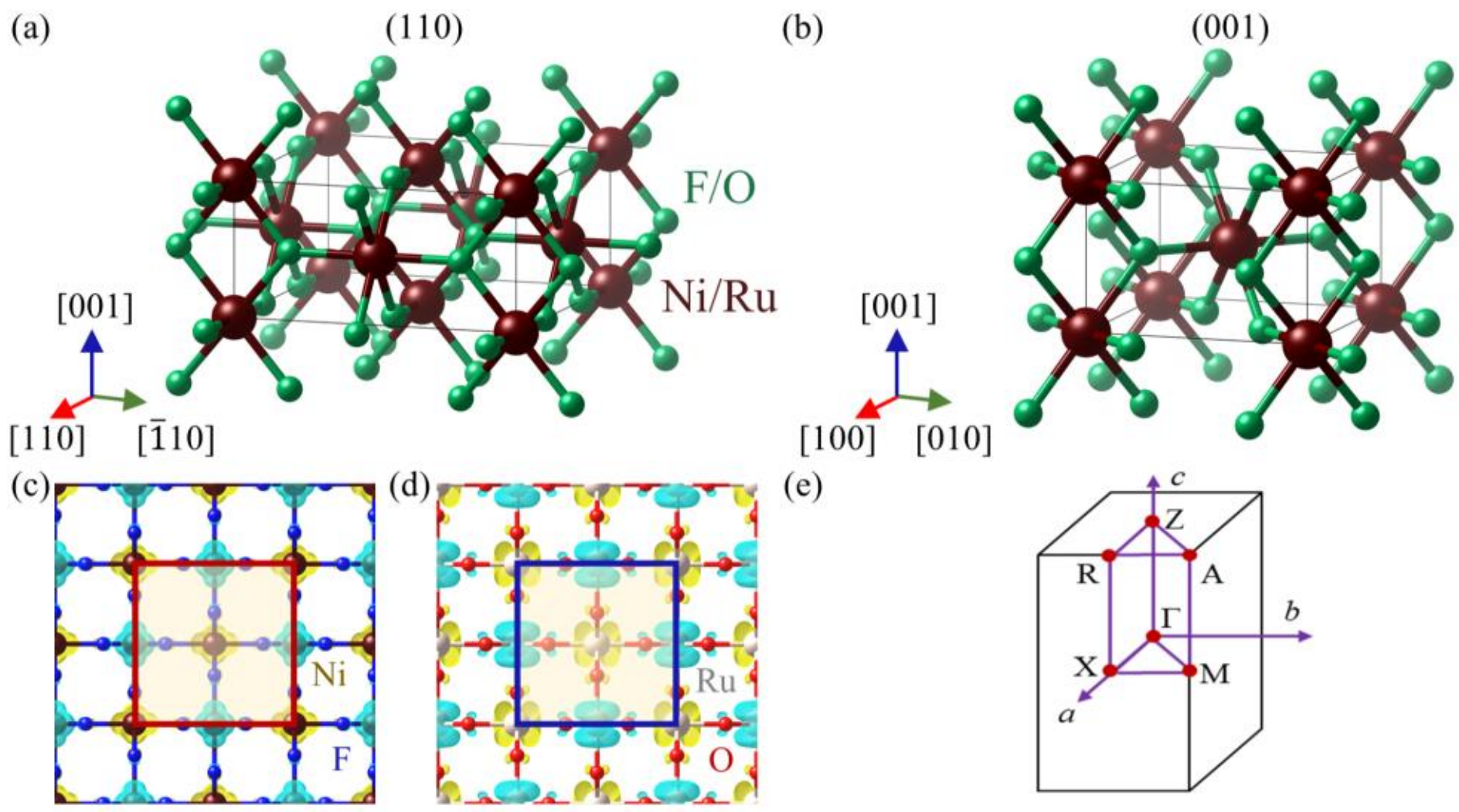

**Fig. 2.** Crystal structures of bulk rutile $NiF_2$ and $RuO_2$ with bulk structures with (110) (a) and (001) (b) crystal faces. Spin-resolved charge density with isosurface of 0.01 e/bohr$^3$ of $NiF_2$ (c) and $RuO_2$ (d). Schematic diagram of high-symmetry points within Brillouin zone (e).

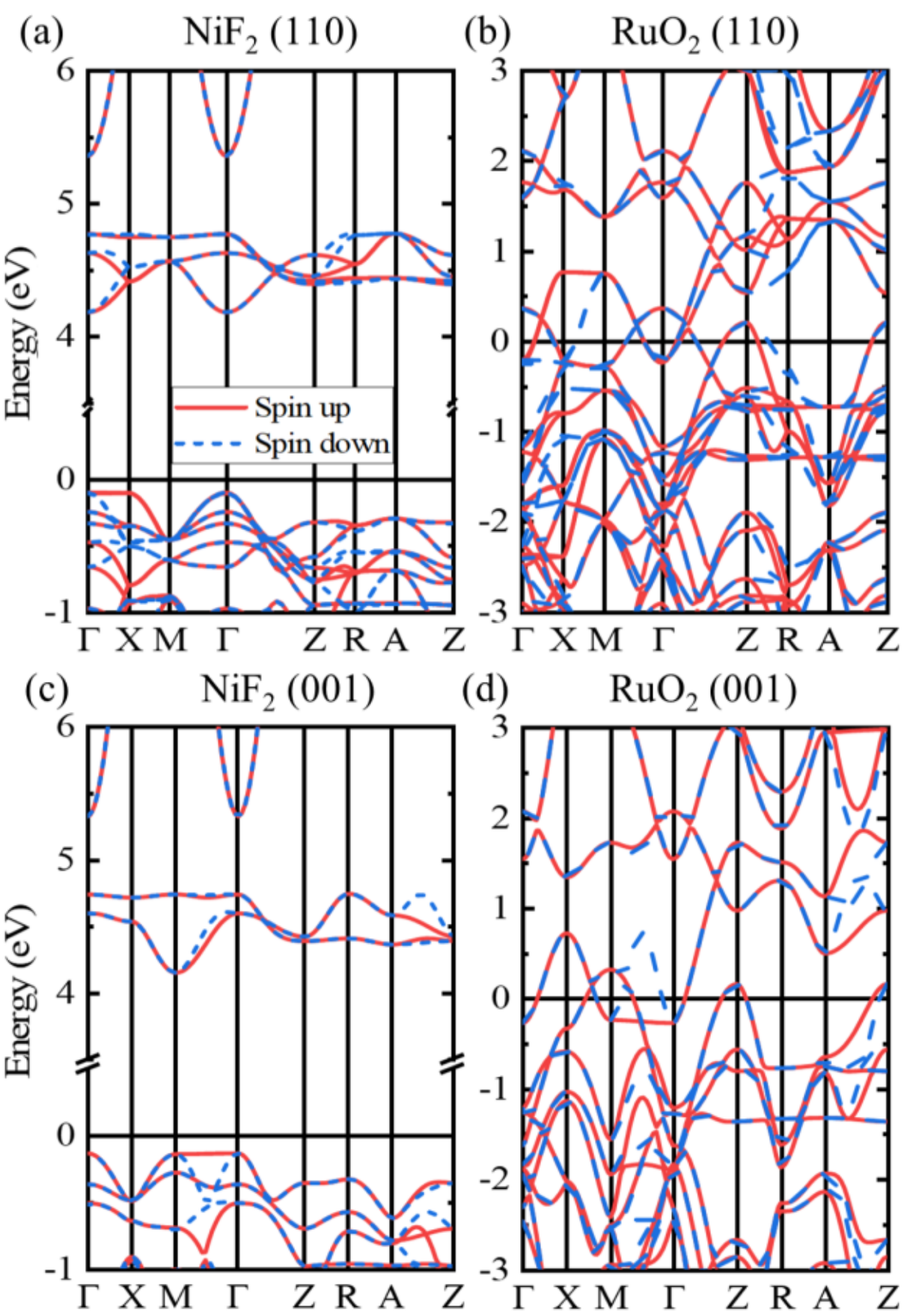

**Fig. 3.** Spin-resolved band structures of $NiF_2$ and $RuO_2$ with bulk structures with (110) (a,b) and (001) (c,d) crystal faces.

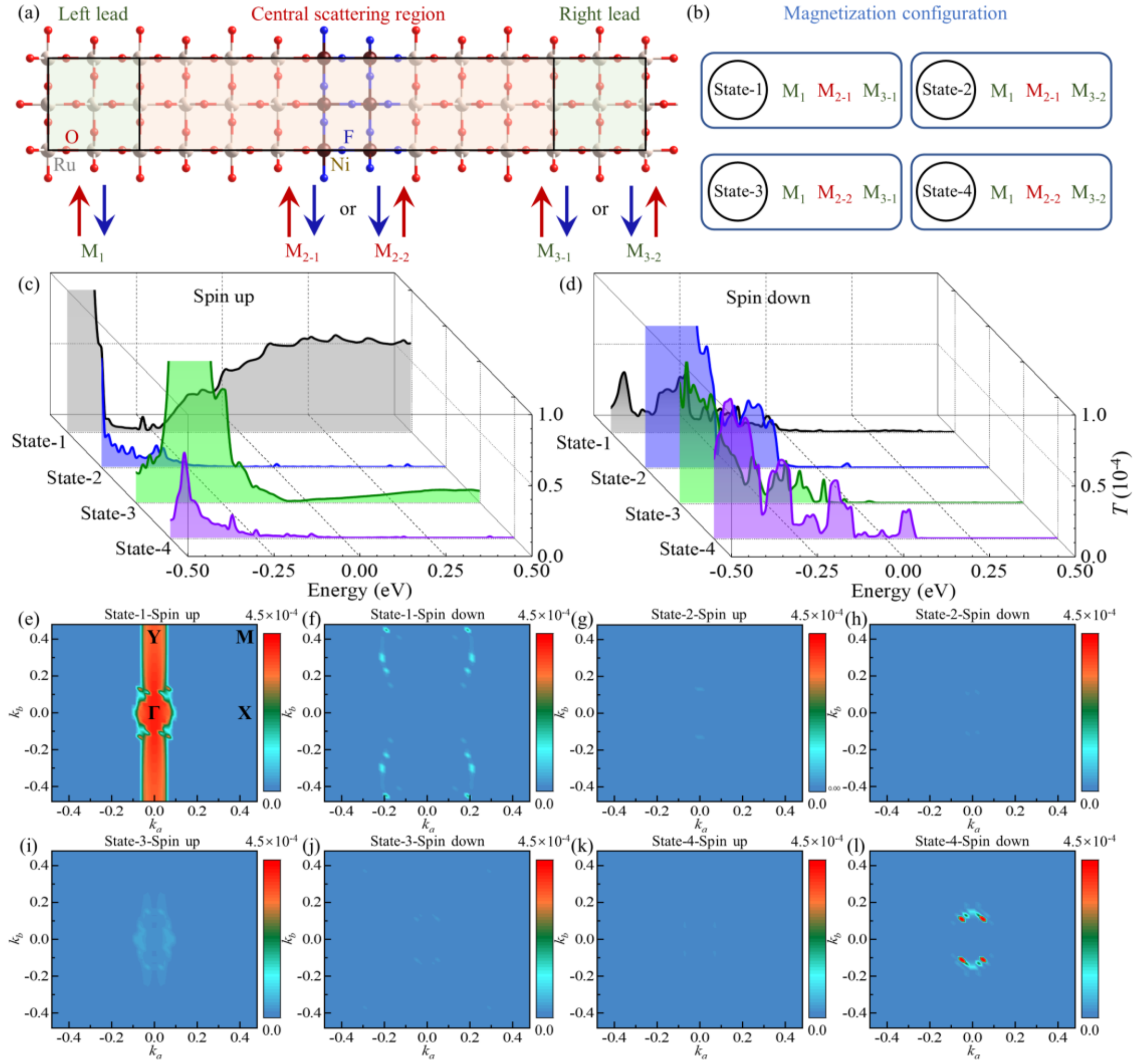

**Fig. 4.** Structural model of the proposed $RuO_2/NiF_2/RuO_2$ AAMTJ (a) and varying magnetization alignments (b). Transmission coefficients as a function of energy in spin-up (c) and spin-down (d) channels. The $\vec{k}_{//}$-resolved transmission spectra in the 2D Brillouin zone (e-l).

**Table 1.** Spin-dependent electron transmission $T_\uparrow$ and $T_\downarrow$ and spin filtering efficiency $\eta$ across $RuO_2/NiF_2/RuO_2$ AAMTJ in the four magnetic configurations.

| | $T_\uparrow$ | $T_\downarrow$ | $T_{tot}$ | $\eta$ (%) |
|---|---|---|---|---|
| State-1 | $4.46 \times 10^{-5}$ | $1.68 \times 10^{-6}$ | $4.62 \times 10^{-5}$ | 93 |
| State-2 | $1.58 \times 10^{-7}$ | $2.34 \times 10^{-7}$ | $3.92 \times 10^{-7}$ | -19 |
| State-3 | $1.65 \times 10^{-6}$ | $1.30 \times 10^{-7}$ | $1.78 \times 10^{-6}$ | 85 |
| State-4 | $1.11 \times 10^{-7}$ | $2.21 \times 10^{-6}$ | $2.32 \times 10^{-6}$ | -90 |

**Table 2.** Multiple tunneling magnetoresistance (TMR) across $RuO_2/NiF_2/RuO_2$ AAMTJ with varying parallel (P) and antiparallel (AP) alignments.

| (P, AP) | (State-1, State-2) | (State-1, State-3) | (State-1, State-4) | (State-3, State-2) | (State-4, State-2) | (State-4, State-3) |
|---|---|---|---|---|---|---|
| TMR (%) | 11,704 | 2,496 | 1,892 | 355 | 493 | 30 |

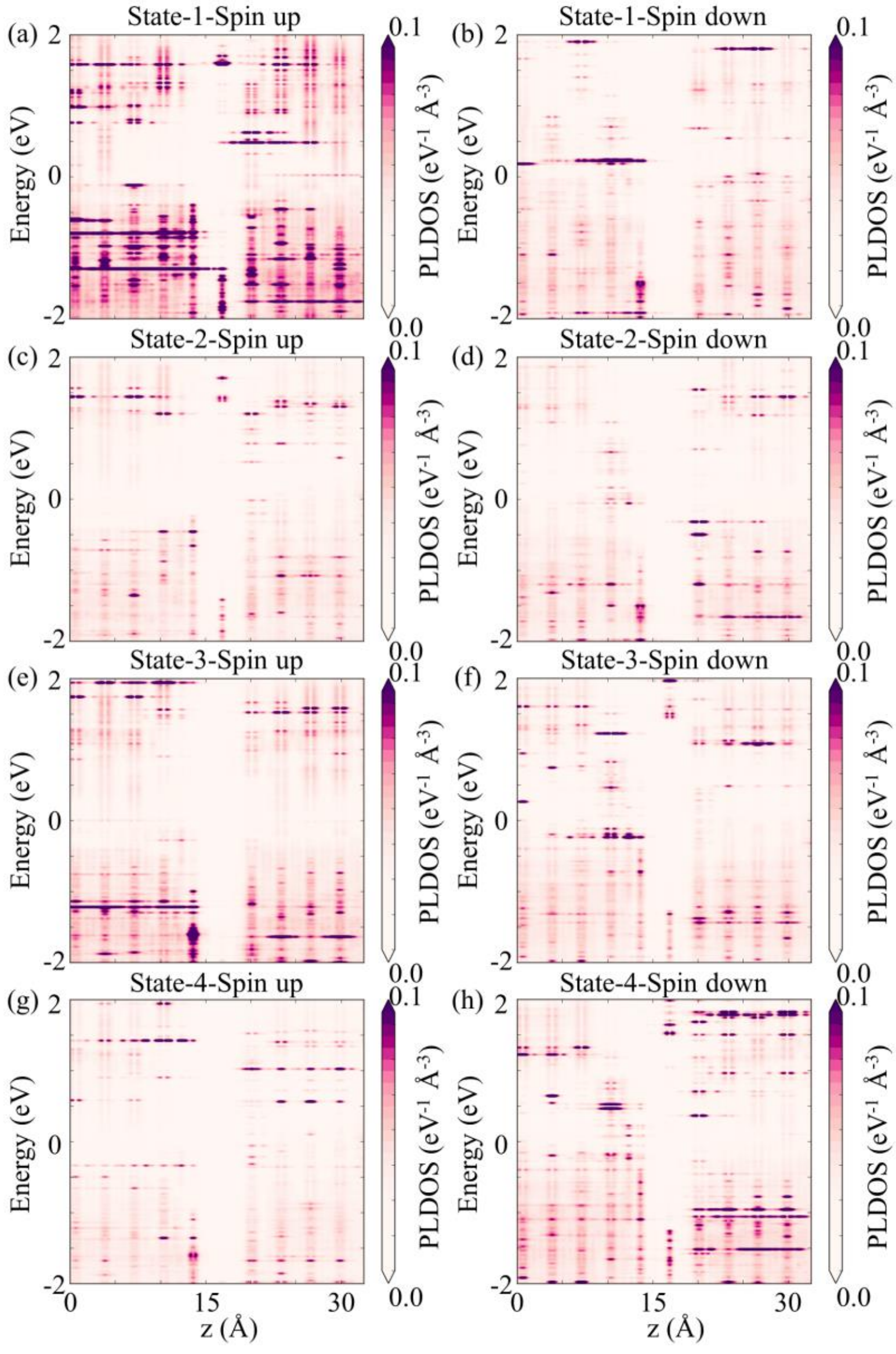

**Fig. 5.** The spin-resolved projected local density of states (PLDOS) across the device of $RuO_2/NiF_2/RuO_2$ AAMTJ with four magnetic configurations. The source of contribution is set as the left $RuO_2$ electrode.

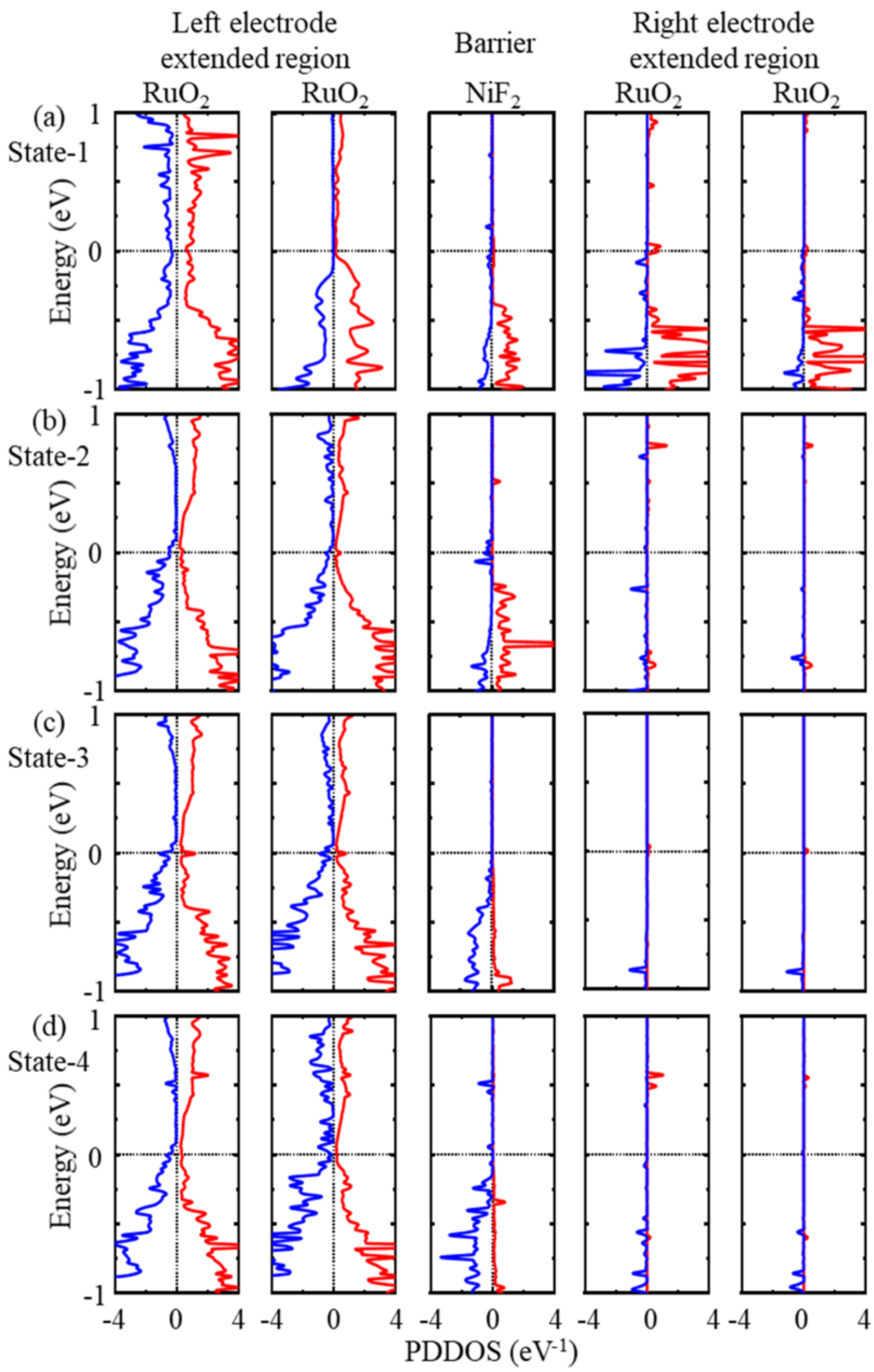

**Fig. 6.** The spin- and layer-resolved projected device density of states (PDDOS) across $RuO_2/NiF_2/RuO_2$ AAMTJ with four magnetic configurations. The source of contribution is set as the left $RuO_2$ electrode.